\documentclass[fleqn,twoside,twocolumn,nofootinbib,showkeys]{revtex4} 
\usepackage[sec,nocpr]{ujp} 

\begin{document}
\title[Analysis of the Stability of Stationary Boundary Friction Modes]
{ANALYSIS OF THE STABILITY OF STATIONARY BOUNDARY FRICTION MODES IN THE
FRAMEWORK OF A SYNERGETIC MODEL}
\author{I.A.~Lyashenko}
\affiliation{Sumy State University}
\address{2, Rimskii-Korsakov Str., Sumy 40007, Ukraine}
\email{nabla04@ukr.net, mtashan@rambler.ru}
 \affiliation{Peter
Gr\"unberg Institut-1}
\address{FZ-J\"ulich, 52425 J\"ulich, Germany}
\author{N.N.~Manko\,}
\affiliation{Sumy State University}
\address{2, Rimskii-Korsakov Str., Sumy 40007, Ukraine}
\email{nabla04@ukr.net, mtashan@rambler.ru}

 \udk{621.891}
\pacs{05.70.Ln; 47.15.gm;\\[-3pt] 62.20.Qp; 68.35.Af;\\[-3pt] 68.60.-p}

\razd{\secix}



\autorcol{I.A.\hspace*{0.7mm}Lyashenko, N.N.\hspace*{0.7mm}Manko}

\setcounter{page}{87}%

\begin{abstract}
A synergetic model describing the state of an ultrathin lubricant layer
squeezed between two atomically smooth solid surfaces operating in
the boundary friction mode has been developed further. To explain
the presence of different operation modes of the system for
various sets of its main parameters, the mathematical analysis of
the synergetic model is carried out. The type of
functioning a tribological system is described in accordance with the stability
character of singular points, and the diagrams distinguishing various
operation modes are obtained. Phase portraits corresponding to
different stability types are plotted for all diagram areas. A
stick-slip mode of motion that is often observed experimentally
is described.
\end{abstract}

\keywords{boundary friction, friction force, shear stresses, strange
attractor, Lorenz system.}

\maketitle

\section{Introduction}

Recently, the processes of boundary friction in nano-sized
tribological systems have become a rather interesting object of
researches for theorists and experimenters
\cite{Persson,Mazyar,Pertsin,Lee,lang1}. Any system, in which the
processes running at the friction of contacting bodies are
essential, can be regarded as tribological\,\footnotemark[1]. The
case of atomically smooth solid surfaces, which move relatively to
each other with the ultrathin layer of a homogeneous lubricant
between them and provided that the distance between the surfaces is
fixed, has not been studied in detail. The application of such
systems for designing a high-precision equipment and devices
\cite{Isrrev} increases the interest in this subject. Note that the
nano-sized systems reveal abnormal properties in comparison with
ordinary macroscopic tribological units. The majority of works
devoted to this topic have a fundamental character
\cite{Popov,Filippov,TribInt1999}. In comparison with bulk
lubricants, the ultrathin lubricant layers are characterized by
different melting and solidification temperatures; they have a
nonmonotonous dependence of the friction force on the velocity,
which follows from the possibility for a lubricant to be in a number
of \mbox{structural states.}

\footnotetext[1]{Below, we consider two elastic contacting solids
under loading and in a state of relative motion, with the thin film
of a lubricant between them.}One of the bright peculiarities
inherent to systems with dry friction is the stick-slip mode of
motion \cite{Pertsin,Capozza,Dry}. This mode has a lot of specific
features and remains unstudied in detail till now despite a
considerable number of theoretical and experimental works (see,
e.g., works \cite{Braun,Benassi} and the references therein). Since
the corresponding experiment in this mode is a complicated task, the
stick-slip mode is often studied with the use of a computer-assisted
simulation \cite{Braun,Benassi}. Work \cite{Benassi} was devoted to
the research of the temperature influence on the emergence of the
boundary friction mode. In work \cite{Capozza}, it was shown
experimentally and theoretically that the stochastic component of
this mode (the time intervals between the beginning moments of the
sticking and slipping modes) is controllable, so that the periods of
the stick-slip mode can be synchronized by changing the magnitude of
a shear force.

Plenty of experimental works formed a basis for the creation of various
theoretical models that describe the boundary friction processes
\cite{Filippov,Popov,Wear1,BraunPRE,VolPerss,TribInt}. One should bear in mind
that even minor changes in both the internal (lubricant type \cite{Lee},
structure of friction surfaces, pressure, and so on) and external
(loading on the surfaces, the shear velocity, the type of tribological
system) parameters can affect the properties of nano-sized systems. For
today, the experimental works provide information concerning such main
properties of a lubricant as its thickness (the number of molecular
layers), temperature, external loading, effective viscosity, elastic and
viscous components of shear stresses, and others \cite{Isrrev}.

In this connection, there appear a considerable number of phenomenological
models. One of them was developed in works \cite{UFJ,JPS,trenieiznos},
where, in the framework of a synergetic representation of boundary friction
and with the use of three differential equations for the stresses, strains,
and temperature in the lubricant layer, the nontrivial behavior of a lubricant
at the relative motion of rubbing surfaces was described. However, the
influence of the parameters in this model on the kinetic modes of dynamic
friction has not been studied in detail. In our previous work
\cite{trenieiznos2013}, we showed that the synergetic model allows the stationary
stick-slip mode of boundary friction to be described in the deterministic
case. This work is a continuation of work \cite{trenieiznos2013}. Here, we will
study the types of stationary state stability and construct the phase diagrams
for various functioning modes of tribological systems.

\section{Basic Equations and Stability Analysis}

The system of basic equations looks like \cite{UFJ,JPS,trenieiznos}
\begin{equation}
\dot\sigma = -\sigma + g\varepsilon, \label{eq1}
\end{equation}\vspace*{-7mm}
\begin{equation}
\tau\dot\varepsilon = -\varepsilon + (T-1)\sigma, \label{eq2}
\end{equation}\vspace*{-7mm}
\begin{equation}
\delta\dot T = (T_e-T) - \sigma\varepsilon + \sigma^2, \label{eq3}
\end{equation}
where $\sigma $ is the shear component of stresses that arise in the
lubricant, $\varepsilon $ the shear component of relative deformations, $T$
the lubricant temperature, and $T_{e}$ the temperature of friction surfaces.
The equations also include the constant $g<1$, which is numerically equal to
the ratio between the shear lubricant modulus $G$ and its characteristic
value $G_{0}$, and the parameters
\begin{equation}
\tau =\tau _{\varepsilon }/\tau _{\sigma },\quad \delta =\tau _{T}/\tau
_{\sigma },  \label{times}
\end{equation}%
where $\tau _{\sigma }$ and $\tau _{\varepsilon }$ are the relaxation times
for the stresses $\sigma $ and the strains $\varepsilon $, respectively; and
the relaxation time for the temperature, $\tau _{T}$, is determined by the
relation
\begin{equation}
\tau _{T}=\rho h^{2}c_{v}/\kappa,
\end{equation}%
where $\rho $ is the lubricant density, $h$ the lubricant
layer thickness, $c_{v}$ the specific heat capacity, and $\kappa $ the thermal
conductivity. The stress $\sigma $, strain $\varepsilon $, temperature $T$,
and time $t$ in the system of equations (\ref{eq1})--(\ref{eq3}) are
reckoned in the corresponding units \cite{UFJ}
\begin{equation}
\begin{array}{l}
\label{eq1d}  \displaystyle \sigma_s = \left(\!\frac{\rho c_v \eta_0
T_c}{\tau_T}\!\right)^{1/2}\!,\quad
\varepsilon_s = \frac{\sigma_s}{G_0},\\[5mm]
 \displaystyle T_s = T_c,\quad t_s = \tau_\sigma,
\end{array}
\end{equation}
where $T_{c}$ is the critical temperature, $G_{0}=\eta _{0}/\tau
_{\varepsilon }$ is a characteristic value of shear modulus, and
$\eta _{0}$ is a characteristic value of shear viscosity. The latter is
expressed in terms of the actual, dimensional viscosity $\eta $ as
follows \cite{oltor}:
\begin{equation}
\eta =\frac{\eta _{0}}{T/T_{c}-1}.
\end{equation}%
Hence, $\eta =\eta _{0}$ at the dimensional temperature $T=$
$=2T_{c}$ or the dimensionless one $T=2$.

In works \cite{UFJ,JPS,trenieiznos}, it was shown that the zero stationary
stresses correspond to a solid-like lubricant structure, whereas, at $\sigma
_{0}\neq 0,$ the lubricant melts and transforms into a liquid-like state. One
of the reasons is that, according to the Stribeck--Hersey diagram
generalized onto the boundary mode \cite{wear}, the growth of viscous
stresses
\begin{equation}
\sigma _{v}=F_{v}/A  \label{f1}
\end{equation}
is accompanied by the growth of the viscous friction force
\begin{equation}
F_{v}=\eta _{\mathrm{eff}}VA/h,  \label{f2}
\end{equation}%
where $V$ is the relative velocity of friction surfaces, $\eta
_{\mathrm{eff}}$ the effective viscosity, and $A$ the contact area.
From Eqs.~(\ref{f1}) and (\ref{f2}), we obtain the following
expression for the velocity:
\begin{equation}
V=\sigma _{v}h/\eta _{\mathrm{eff}}.  \label{V_block}
\end{equation}%
Since the stress $\sigma $ in the proposed model is a sum of the
viscous and elastic components \cite{UFJ}, and the viscous stresses
prevail in the liquid-like lubricant layer, the velocity of motion
of shear surfaces increases with $\sigma,$ which corresponds to
the kinetic slipping mode and the liquid-like lubricant structure.
At $\sigma =0$, the friction surfaces do not move, which corresponds
to their \textquotedblleft sticking\textquotedblright\ owing to the
solidification of the layer between the surfaces. Those conclusions
are confirmed both theoretically \cite{Popov} and
\mbox{experimentally \cite{lang1}.}

The initial system of equations (\ref{eq1})--(\ref{eq3}) has a general
character; it does not make allowance for the properties of a specific
tribological system. Therefore, it can describe the features of the boundary
mode in tribological systems of various types. Let us consider two of them,
which are the most widespread (Fig.~1).

Figure~1,~\textit{a} schematically illustrates a system consisting
of a spring with the stiffness constant $k$ and connected with a
block of mass $M$ located on a smooth motionless surface; a
lubricant layer of thickness $h$ separates the block and the
surface. An additional loading $L$ is applied normally to the block.
The free end of the spring moves at the velocity $V_{0}$. The system
depicted in Fig.~1,~\textit{b} is composed of a spring connected to
a block on rollers (the rolling friction of the rollers is
neglected). Another block is located on the surface of the first
block. The velocity of the motion of the second block, $V_{0}$,
changes periodically \cite{JtfL2,TribInt1999}. Provided that there
is an ultrathin lubricant layer between blocks' surfaces, the
motion of the upper block stimulates the lower one to move, with the
time dependence of the motion velocity for the lower block, $V(t)$,
depending substantially on the friction mode.

Note that the system exhibited in Fig.~1,~\textit{a} was studied
both experimentally and in the framework of two thermodynamic models
\cite{TribInt,JtfL3} basing on the Landau theory of phase
transitions, as well as in the framework of a stochastic model,
which takes the interaction between the friction surfaces into
consideration \cite{Filippov}. In contrast to work \cite{JtfL3}, the
model analyzed in work \cite{TribInt} makes allowance for the
influence of the external loading $L$ explicitly. The installation
depicted in Fig.~1,~\textit{b} was studied experimentally in work
\cite{TribInt1999} and analyzed in the framework of a thermodynamic
model in work \cite{JtfL2}.

\begin{figure}%
\vskip1mm
\includegraphics[width=\column]{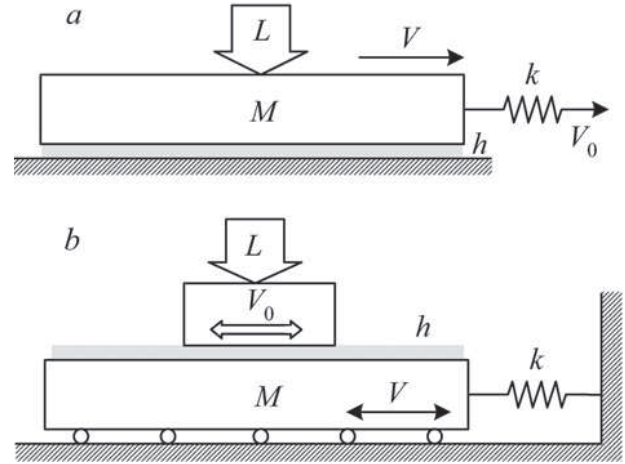}
\vskip-3mm\caption{Mechanical analogs of tribological systems of two
types  }
\end{figure}

The solution of the system of differential equations
(\ref{eq1})--(\ref{eq3}) in a vicinity of the stationary state is sought in the
form
\begin{equation}
\sigma = \sigma_0+\alpha e^{\lambda t}, \label{eq4}
\end{equation}\vspace*{-7mm}
\begin{equation}
\varepsilon = \varepsilon_0+\beta e^{\lambda t}, \label{eq5}
\end{equation}\vspace*{-7mm}
\begin{equation}
T = T_0+\gamma e^{\lambda t}, \label{eq6}
\end{equation}
where $\lambda $ is an unknown increment, the parameters $\sigma _{0}$,
$\varepsilon _{0}$, and $T_{0}$ correspond to the stationary state, and the
amplitudes $\alpha $, $\beta $, and $\gamma $ characterize small deviations
from this state. Substituting Eqs.~(\ref{eq4})--(\ref{eq6}) into system
(\ref{eq1})--(\ref{eq3}), we arrive at a system of algebraic equations
\begin{equation}
\alpha\lambda e^{ \lambda t} = - \sigma_0-\alpha e^{\lambda t}
 + g\left(\varepsilon_0+\beta e^{\lambda t}\right)\!, \label{eq7}
\end{equation}\vspace*{-7mm}
\[
\tau \beta \lambda e^{\lambda t} = - \varepsilon_0-\beta e^{\lambda
t} +
\]\vspace*{-7mm}
\begin{equation}
+\left(T_0+\gamma e^{\lambda t} -1\right)\left(\sigma_0+\alpha
e^{\lambda t}\right)\!, \label{eq8}
\end{equation}\vspace*{-7mm}
\[
\gamma \lambda  e^{ \lambda t} = T_e-T_0-\gamma e^{\lambda t} -
\left(\sigma_0+\alpha e^{\lambda t}\right) \left(\varepsilon_0+\beta
e^{\lambda t}\right) +
\]\vspace*{-7mm}
\begin{equation}
+\left(\sigma_0+\alpha e^{\lambda t}\right)^2\!. \label{eq9}
\end{equation}
The roots of this system at $(\alpha,\beta,\gamma )\rightarrow 0$ are
stationary values. The analysis of Eqs.~(\ref{eq7})--(\ref{eq9}) allows us
to define the critical temperature $T_{c0}$. Namely, if the temperature of
friction surfaces, $T_{e}$, is lower than
\begin{equation}
T_{c0}=1+g^{-1},  \label{eqCrit}
\end{equation}
the stationary values
\begin{equation}
\sigma _{0}=\varepsilon _{0}=0,\quad T_{0}=T_{e},  \label{eq13_1}
\end{equation}%
are realized, whereas at $T_{e}>T_{c0}$, either of two sets, ($\sigma
_{0}^{+}$,$\varepsilon _{0}^{+}$,$T_{0}$) or ($\sigma _{0}^{-}$,$\varepsilon
_{0}^{-}$,$T_{0}$), where
\begin{equation}
\begin{array}{l}
\displaystyle\sigma_0^\pm = \pm \sqrt \frac {g T_e -1-g}{1-g},\\[3mm]
\displaystyle\varepsilon_0^\pm = \pm \frac {1}{g} \sqrt \frac {g T_e
-1-g}{1-g},~~~  T_0 = 1 + g^{-1}, \label{eq13}
\end{array}
\end{equation}
is established depending on the initial conditions; the both
correspond to the liquid-like friction mode. Hence, at
$T_{e}>T_{c0}$\ (see Eq.~(\ref{eqCrit})), the lubricant melts
\cite{UFJ,JPS,trenieiznos}.

In a first order of smallness with respect to the parameters $\alpha $,
$\beta $, and $\gamma $, the system of equations (\ref{eq7})--(\ref{eq9})
reads
\begin{equation}
(\lambda+1)\alpha - g\beta = 0, \label{eq17}
\end{equation}\vspace*{-7mm}
\begin{equation}
(1-T_0)\alpha + (\tau\lambda+1)\beta-\sigma_0\gamma  = 0,
\label{eq18}
\end{equation}\vspace*{-7mm}
\begin{equation}
(2\sigma_0-\varepsilon_0)\alpha + \sigma_0\beta+(\delta\alpha  +
1)\gamma  = 0, \label{eq19}
\end{equation}
where $\sigma _{0}$, $\varepsilon _{0}$, and $T_{0}$ are the stationary
values. Therefore, the solutions of the system for $T_{e}$ above and below the
critical value (\ref{eqCrit}) are different. Let us consider firstly the
temperature interval ${T_{e}<T_{c0}}$ (the solid-like lubricant). In this
case, the stationary state (\ref{eq13_1}) is realized. A nontrivial solution
of the system of equations (\ref{eq17})--(\ref{eq19}) exists if the
determinant equals zero,
\begin{equation}
\left( \delta \lambda +1\right) \left[ \tau \lambda ^{2}+(\tau +1)\lambda
+g(1-T_{e})+1\right] =0.  \label{eq19_1}
\end{equation}
Equation (\ref{eq19_1}) has the following three roots:
\begin{equation}
\begin{array}{l}
\displaystyle \lambda_{1,2} =
\frac{-(\tau+1)\pm\sqrt{\tau^2+(4gT_e-4g-2)\tau+1}}{2\tau},\\[3mm]
\displaystyle \lambda_3 = -1 / \delta, \label{eq13_3}
\end{array}\!\!\!\!
\end{equation}
so that the root $\lambda _{3}$ is always negative. Concerning two
other roots, which are complex-conjugate with respect to each other,
their real part is also always negative. The form of a root
governs the type of stability of a stationary point (see below).

In the liquid-like state of the lubricant ($T_{e}>T_{c0}$), the stationary
values (\ref{eq13}) are realized. In this case, substituting those values
into system (\ref{eq17})--(\ref{eq19}) and equating the corresponding
determinant to zero, we obtain the cubic equation
 \begin{equation}
\lambda ^{3}+A\lambda ^{2}+B\lambda +C=0  \label{eq20}
\end{equation}
with the coefficients
\begin{equation}
\begin{array}{l}
\displaystyle A = 1+ \tau^{-1} + \delta^{-1},\\[3mm]
\displaystyle B = \delta^{-1} + \frac{g(2-T_e)}{\tau\delta (g-1)},\\[3mm]
\displaystyle C = \frac{2\left(gT_e-1-g\right)}{\tau\delta}.
\label{eq23}
\end{array}
\end{equation}
We now write the discriminant of Eq. \eqref{eq20}:
\begin{equation}
\Delta = -4A^3C + A^2B^2 - 4B^3 + 18ABC - 27C^2. \label{eq20_D}
\end{equation}
If $\Delta >0$, Eq.~(\ref{eq20}) has three different real-valued
roots. The equality $\Delta =0$ means that two real roots coincide.
At $\Delta <0$, we have one real and two complex-conjugate roots.

To analyze eigenvalues (\ref{eq13_3}) and the solution of
Eq.~(\ref{eq20}), let us plot a diagram, where the regions would be
characterized by different $\lambda _{i}$-values (Fig.~2). The curve
separating regions 1 and 5 in Fig.~2,~\textit{a} can be found by
analyzing solution (\ref{eq13_3}), because $T_{e}<T_{c0}$ in this
region. In all figures, regions~1 and 2 are separated by condition
(\ref{eqCrit}) (dashed curves). The analysis of discriminant
(\ref{eq20_D}) (the equation $\Delta =0$) allows us to plot the
curve that separates region~2 from region~3. Let us introduce the
additional condition $AB=C$ (see Eq.~(\ref{eq23})), which allows us
to plot the curve, where the system loses its
stability.\footnote[2]{In this case, Eq.~(\ref{eq20}) can be written
down in the form $(\lambda -$ $-a)(\lambda -bI)(\lambda +bI)=\lambda
^{3}-a\lambda ^{2}+b^{2}\lambda -ab^{2}$, where $a$ and $b$ are
real numbers, and $I$ is the imaginary unity. The given condition can
be expressed in the form $AB=C$ with the coefficients $A=-a$,
$B=b^{2}$, and $C=-ab^{2}$.} In Fig.~2,~\textit{a}, it is a curve
that separates regions~3 and 4. The presence of the complex part and
the sign of $\lambda _{i}$ determine the type of a singular point for
the given set of parameters $T_{e}$, $g$, $\tau $, and $\delta $. In
other words, the type of a stationary mode at fixed model parameters
can be sometimes determined beforehand, if the $\lambda _{i}$-values
are known. In Fig.~2,~\textit{a}, there exist five regions for the
selected set of parameters. In regions~1 and 5, the singularity
point has coordinates (\ref{eq13_1}), and the corresponding
eigenvalues are determined by Eq.~(\ref{eq13_3}). Regions~2, 3, and
4 were plotted for the parameter sets, at which $T_{e}>T_{c0}$;
the singularity points (\ref{eq13}) are realized in them, and the
eigenvalues are determined as the roots of Eq.~(\ref{eq20}).

In Table, the roots of Eq.~(\ref{eq20}) are listed, as well as
eigenvalues (\ref{eq13_3}) for each of five regions (the
corresponding points are marked by diamonds in Fig.~2). In
Fig.~2,~\textit{a,} three types of singular points are
distinguished: stable nodes (regions~1 and 2), stable foci
(regions~3 and 5), and a saddle-focus (region~4). Regions with the
same stability type correspond to different singular points (the
stationary state before (Eq.~(\ref{eq13_1})) and after
(Eq.~(\ref{eq13})) melting). Region~2 (the stable node) in
Fig.~2,\textit{a} is very narrow, but it separates regions~1 and 3
within the whole interval of selected values. For illustration,
regions~2 and 5 are shown on a large scale in the inset in
Fig.~2,~\textit{a} (the $T_{e}$-axis contains a break).
Figures~2~\textit{b} and \textit{c} are plotted in different
coordinates, but include the same regions.

The system of equations (\ref{eq1})--(\ref{eq3}) is reduced to a
differential equation of the third order \cite{trenieiznos2013},
\[
\dddot\sigma -
\ddot\sigma\left(\!\frac{\dot\sigma}{\sigma}-1-\frac{1}{\tau}
-\frac{1}{\delta}\!\right) -
\]\vspace*{-7mm}
\[
-\,\frac{\dot\sigma}{\tau}\left(\!\frac{\dot\sigma\left(\tau+1\!\right)}{\sigma}
-\frac{\sigma^2+1+\tau}{\delta}\right) -
\]\vspace*{-7mm}
\begin{equation}
-\,\frac{\sigma\left(g\left(T_e+\sigma^2-1\right)-\sigma^2
-1\right)}{\tau\delta} = 0. \label{eq_main}
\end{equation}

\begin{table}[b]
\noindent\caption{Roots of Eqs.~(\ref{eq19_1}) and (\ref{eq20})\\
for various diagram regions shown in Fig.~2}\vskip3mm\tabcolsep2.0pt

\noindent{\footnotesize\begin{tabular}{|c|c|c|c|c|l|l|}
 \hline \multicolumn{1}{|c}
{\rule{0pt}{5mm}No.} & \multicolumn{1}{|c}{$\tau$}&
\multicolumn{1}{|c}{$\delta$}& \multicolumn{1}{|c}{$g$}&
\multicolumn{1}{|c}{$T_e$}& \multicolumn{1}{|c}{$\lambda_{1,2,3}$}&
\multicolumn{1}{|c|}{Point type}\\[2mm]%
\hline%
\rule{0pt}{5mm}$1'$&$3$& $0.6$ & $0.2$ & $4$ & $-1.6667, -1.2244, -0.1089$ &stable node \\
$2'$&$20$& $3$ & $0.5$ & $4$ & $-1.0489, -0.2771, -0.0573$ &stable node \\
$3'$&$3$& $30$ & $0.2$ & $30$ & $-1.3431, -0.0118\pm 0.2816I$ &stable focus\\
$4'$&$3$& $30$ & $0.2$ & $120$ & $-1.3725, 0.0029\pm 0.6076I$ &saddle-focus \\
$5'$&$3$& $30$ & $0.8$ & $0.5$ & $-0.0333, -0.6667\pm 0.1491I$
&stable focus
\\[2mm]
\hline
\end{tabular}\label{tab1}}
\end{table}

\begin{figure}%
\vskip1mm
\includegraphics[width=5.5cm]{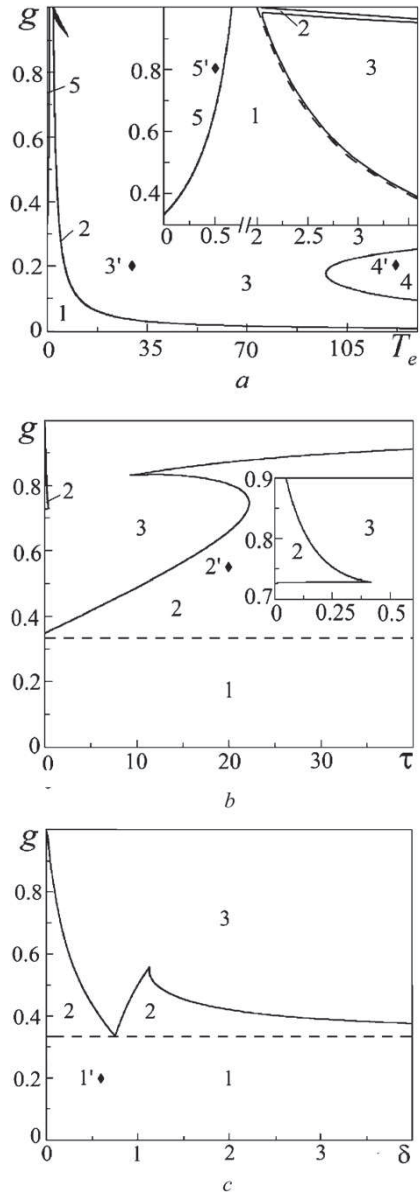}
\vskip-3mm\caption{Diagrams of stability types for singular points:
(\textit{a})~$\tau =$ $=3$, $\delta=30$; (\textit{b})~$T_{e}=4$,
$\delta=3$; (\textit{c})~$\tau=3$, $T_{e}=4$  }
\end{figure}

\begin{figure}%
\vskip1mm
\includegraphics[width=6.5cm]{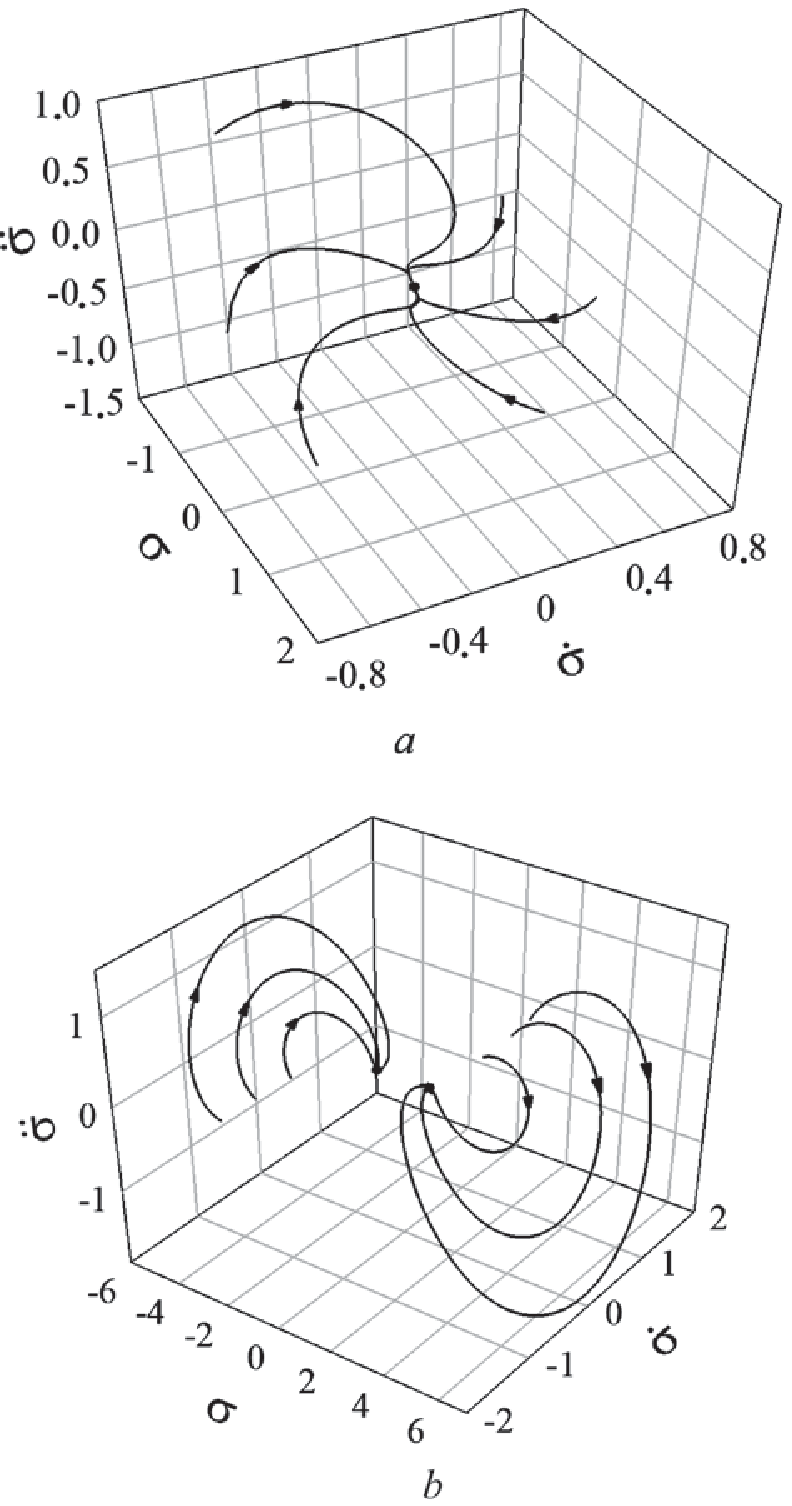}
\vskip-3mm\caption{Phase portraits of Eq.~(\ref{eq_main})
corresponding to the parameters of points~1$^{\prime}$ (\textit{a})
and 2$^{\prime}$ (\textit{b})  }
\end{figure}

\noindent In this equation, the stresses $\sigma $, in accordance
with Eq.~(\ref{V_block}), are proportional to the velocity of
relative motion of the friction surfaces, $V$; therefore, $\dot{\sigma}$
is the acceleration. Let us plot the phase portraits of the system
for all regions in the diagram. For this purpose, we should
numerically solve Eq.~(\ref{eq_main}) with the use of the Runge--Kutta
method of the fourth order and plot figures for the diagram regions
with the parameters taken from Table (for all five points). The
stable node is shown in Fig.~3. Figure~3,~\textit{a} corresponds to
the temperature $T_{e}<T_{c0}$, at which one stationary point
(\ref{eq13_1}) is realized (region~1). This case corresponds to the
lubricant solidification in time. Figure~3,~\textit{b} corresponds
to the temperature $T_{e}>T_{c0}$. At this temperature, two
symmetric singular points \ref{eq13} are realized (region~2), which
correspond to the liquid-like lubricant. From the shape of
trajectories in Fig.~3,~\textit{a}, it follows that, before the
stationary state is established (i.e. the system stops), an
aperiodic transient mode takes place, in which the stresses relax to
the stable value $\sigma =0$, which corresponds to the solid-like
lubricant structure. In this case, the mobile block shown in Fig.~1
stops in due course. It can be realized in the case shown in
Fig.~1,~\textit{a} at the velocity $V_{0}=0$. This situation
also describes the behavior of the system shown in
Fig.~1,~\textit{b} at $V_{0}=0$, when the lower block is initially
not in the equilibrium state, i.e. the spring is either compressed
or tensed. In Fig.~3,~\textit{b}, the both singular points are
equivalent and, as well as in Fig.~3,~\textit{a}, correspond to
stable nodes. However, for all selected initial conditions, a motion
at a constant velocity is eventually established. This situation
corresponds to the tribological system exhibited in
Fig.~1,~\textit{a} at $V_{0}\neq 0$. If $\sigma <0$, the friction
surface moves with a negative velocity, i.e. in the opposite
direction (the reverse motion).

\begin{figure}
\vskip1mm
\includegraphics[width=6.6cm]{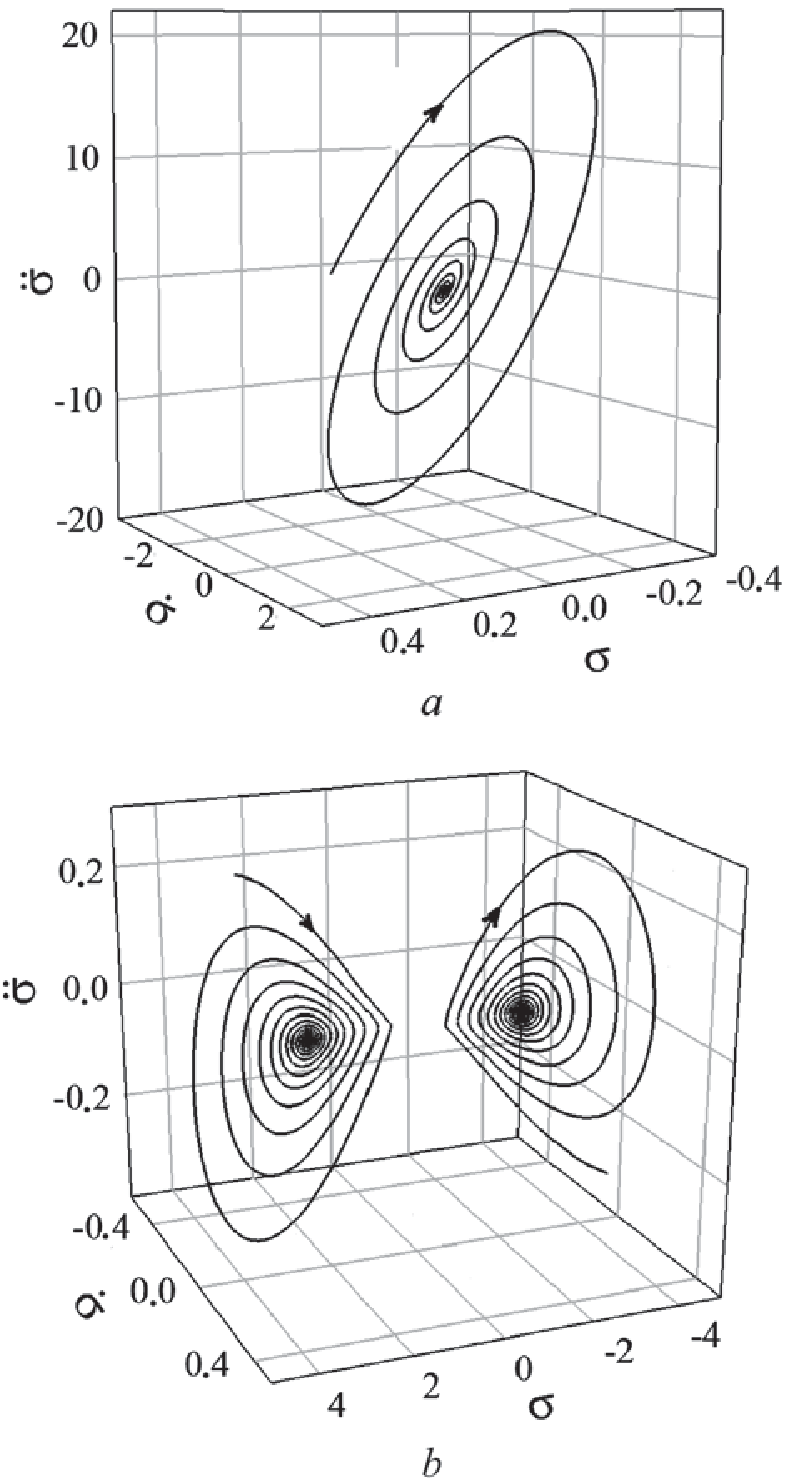}
\vskip-3mm\caption{Phase portraits of Eq.~(\ref{eq_main})
corresponding to the parameters of points~5$^{\prime}$ (\textit{a})
and 3$^{\prime}$ (\textit{b})  }
\end{figure}

A convergence in the form of stable focus is also represented by two
regions, 3 and 5, where the singular points (\ref{eq13}) and
(\ref{eq13_1}), respectively, are realized. At the parameters of
region~5 (Fig.~4,~\textit{a}), the temperature is lower than the
melting one; therefore, the trajectories in the phase portrait
converge to the singular point (\ref{eq13_1}). The phase portrait
depicted in Fig.~4,~\textit{b} has the parameters of region~3;
therefore, it corresponds to the melted lubricant (two nonzero
singular points are realized). For this type of stability, long-term
oscillations take place in the system concerned before a motion with
a constant velocity is established (Fig.~4,~\textit{b}) or the
system stops (Fig.~4,~\textit{a}). Moreover, at some parameters, the
trajectories do not converge to singular points, but a chaotic mode
is realized \cite{trenieiznos2013}. In the latter case, to elucidate
the features of this mode, an additional analysis is required. In
the case shown in Fig.~4,~\textit{a}, the mobile block (see
Fig.1,\textit{a}) also eventually stops, as in Fig.3,~\textit{a}.
This situation also describes the behavior of the system shown in
Fig.~1,~\textit{b} at $V_{0}=0$, when the lower block is initially
not in the equilibrium state, i.e. the spring is either compressed
or tensed. In due course, when a motion with a constant velocity is
established (Fig.~4,~\textit{b}), the situation corresponds to the
tribological system shown in Fig.~1,~\textit{a} at $V_{0}\neq 0$.

\begin{figure}
\vskip1mm
\includegraphics[width=6.5cm]{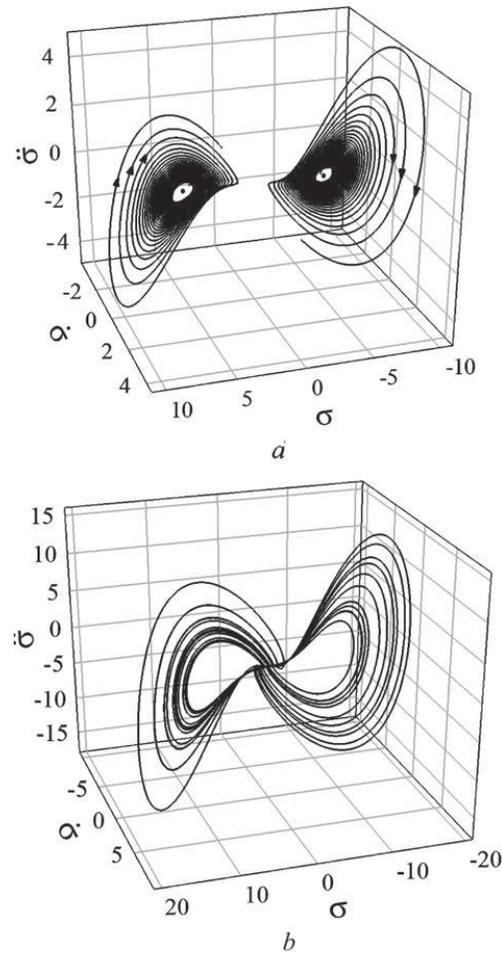}
\vskip-3mm\caption{Phase portraits of Eq.~(\ref{eq_main})
corresponding to the parameters of point~4$^{\prime}$:
(\textit{a})~near the singular points and (\textit{b})~in the
stationary mode  }
\end{figure}

A convergence of the saddle-focus type can reveal itself as a saddle
or a focus, depending on the parameter set selected from the region.
The phase portrait exhibited in Fig.~5,~\textit{a} is characterized
by two symmetric singular points representing unstable foci. The
considered diagram region corresponds to condition (\ref{eq13}). In
this region, the motion with a constant velocity cannot be established
because the system enters the chaotic mode of functioning. Hence,
the stationary mode is accompanied by permanent phase transitions
between the solid-like and liquid-like lubricant states. This mode
is not periodic in time but is a strange attractor, i.e. it realizes
a deterministic chaotic mode in the system \cite{trenieiznos2013}.
Since the velocity of the friction block motion permanently changes
its sign, the considered situation describes the behavior of the
system shown in Fig.~1,~\textit{b}, when the motion direction
changes under the action of an external influence. However, at large
$V_{0}$-values, the examined reverse mode can also be realized in
the systems, whose analog is shown in Fig.~1,~\textit{a}. This
occurs because, when the free end of the spring moves at a large
velocity $V_{0}$, the spring itself has enough time to be strongly
stretched within the time interval of the surface \textquotedblleft
sticking\textquotedblright\ (i.e. when $\sigma =0$ and the lubricant
is solid-like). On the other hand, as the lubricant melts, the block
slides at a large distance owing to a large magnitude of elastic
force $k\Delta x$ ($\Delta x$ is the string elongation), the spring
becomes compressed at that, and the direction of the elastic force
changes. As a result, the block may move in the opposite direction
for some time \cite{trenieiznos2013}. Figure~5,~\textit{b} (the
stationary mode at the parameters of Fig.~5,~\textit{a}) exhibits a
Lorenz butterfly, which in well-known in the chaos theory
\cite{Loskutov,Lorenz}. Hence, Fig.~2 reveals the existence of
essentially different friction modes. At the same time, this figure
also demonstrates that such modes can emerge in tribological systems
of various types.

\section{Conclusions}

In this work, we made a further research of the synergetic model that
describes the state of an ultrathin lubricant layer squeezed between
atomically smooth solid surfaces in the course of boundary friction. It is
shown that this model can describe the behavior of tribological systems of
various types. A number of stationary solutions, which correspond
to the dry and liquid friction modes, are found, as well as the ranges of model
parameters, at which that or another mode of tribological system functioning
is established. In addition, different modes characterized by one of three
types of convergence (a stable node, a stable focus, or a saddle-focus) are
distinguished. For each mode, the phase portrait is plotted, and the
behavior of tribological system is described. The growth of the friction
surface temperature is found to result in an enhancement of the stochasticity
in the system. If the temperature exceeds the critical value, the mode of
functioning of the system is described by the Lorenz attractor. In a wide range of
parameters, the reverse motion of the friction surfaces is realized. The results
obtained qualitatively coincide with known experimental data.

\vskip3mm

\textit{The work was sponsored by the State Fund for Fundamental
Researches of Ukraine in the framework of the grant of the President
of Ukraine GP/F44/010 \textquotedblleft Phenomenological theory of
boundary friction in tribological nanosystems\textquotedblright,
No.~0112U007318. Some results were obtained under the support of the
Ministry of Education and Science, Youth and Sport of Ukraine in the
framework of the project \textquotedblleft Simulation of friction of
metallic nanoparticles and boundary films of liquids interacting
with atomically smooth surfaces\textquotedblright, No.~0112U001380.
The work was supported by the KMU grant. The research was carried
out during the stay of I.A.L. at the Forschungszentrum J\"{u}lich
(Germany) on the invitation of B.N.J.~Persson.}

\rezume{%
Я.О.~Ляшенко, Н.М.~Манько}{АНАЛІЗ СТІЙКОСТІ СТАЦІОНАРНИХ\\ РЕЖИМІВ
МЕЖОВОГО ТЕРТЯ В РАМКАХ \\СИНЕРГЕТИЧНОЇ МОДЕЛІ} {У даній роботі
проведено подальший розвиток синергетичної моделі, яка описує стан
ультратонкого шару мастила, що затиснутий між двома
атомарно-гладкими твердими поверхнями, які працюють у режимі
межового тертя. Проведено математичний аналіз синергетичної моделі з
метою пояснення виникнення різних режимів роботи системи при зміні
головних параметрів. Тип функціонування трибологічної системи
описаний у відповідності з характером  стійкості особливих точок.
Отримано діаграми, на яких виділено різні режими роботи. Для всіх
областей діаграм побудовано фазові портрети, що відповідають різним
типам стійкості. Описано переривчастий рух, що часто зустрічається в
експериментах.}

\end{document}